\documentclass[referee,usegraphicx]{mn2e}
\newif\ifAMStwofonts
\AMStwofontstrue

\ifCUPmtlplainloaded \else
  \ifAMStwofonts \else % If no AMS fonts

  \fi
\fi
\title{On the effects of the stellar magnetic field on the structure of
T Tauri accretion discs}
\author[Ultchin et al.]
       {Y. Ultchin,$^1$ O. Regev$^1$ and G. A. Wynn$^2$ \\
        $^1$ Physics Department, Technion - Israel Institute of Technology
        , Haifa 32000, Israel\\
        $^2$ Astronomy Group, Leicester University, University Road, Leicester LE1 7RH}
\date{Accepted -------. Received -------; in original form -------.}

\pagerange{\pageref{firstpage}--\pageref{lastpage}} \pubyear{2001}

\begin{document}

\maketitle

\label{firstpage}

\begin{abstract}
The structure of accretion discs around magnetic T Tauri stars is
numerically calculated using a particle hydrodynamical code, in
which magnetic interaction is included in the framework of King's
dimagnetic blob accretion model. Setting up the calculation so as
to simulate the density structure of a quasi-steady disc in the
equatorial plane of a T Tauri star, we find that the central
star's magnetic field typically produces a central hole in the
disc and spreads out the surface density distribution. We argue
that this result suggets a promising mechanism for explaining the
unusual flatness (IR excess) of T Tauri accretion disc spectra.
\end{abstract}


\section{Introduction}

It is now quite widely accepted that in the formation process of
low-mass ($M \la 2 M_\odot$) stars, a disc accretion phase is
typically present (see Hartmann 1998 for an extensive review and
references). In particular, the classical T Tauri stars (CTTS) are
thought to be still surrounded by discs and the emission
properties of some typical objects suggest that these discs are
not merely ``passive" dust bodies, irradiated by the central star,
but are rather bona-fide energy producing, viscous accretion
discs, similar to the ones present in mass transferring close
binary systems, and whose modeling dates back to the classical
works of Shakura \& Sunyaev (1973) and Lynden-Bell \& Pringle
(1974). Because of their ubiquity and of some interesting
theoretical challenges they pose, accretion discs have remained at
the centre stage of astrophysical research (see e.g. Frank, King
\& Raine 1992; Papaloizou \& Lin 1995; Lin \& Papaloizou 1996 for
reviews).

One particular issue in this context is the fact that in many CTTS
the measured spectrum presents an unusually high emission in the
infra red (called IR excess), far more than expected from disc
reprocessing of stellar light or viscous heating of the disc (see
Bertout 1989, Beckwith et al. 1990, Hartmann et al. 1994). Our
wish to investigate this problem, as well as some other facets of
CTTS discs, has been the main motivation for the work reported on
in this paper. It consists of setting up a numerical calculation
of the properties of an accretion disc around magnetic CTTS, using
a scheme of magnetic interaction, originally proposed in the
context of diamagnetic accretion by King \shortcite{king1}.

This description of the magnetic interaction assumes that as
material moves through the magnetosphere it interacts with the
local magnetic field via a velocity dependent acceleration (force
per unit mass, $f_{\rm mag}$) of the form:
 \begin{equation}
 f_{\rm mag} = - K \, [{\bf u} - {\bf u}_f]_\bot,
 \label{general}
\end{equation}
where ${\bf u}$ and ${\bf u}_f$ are the velocities of the material
and magnetic field lines respectively, $K$ is a suitable "magnetic
drag" coefficient (see below) and the suffix $\perp$ refers to the
vector component perpendicular to the field line. The magnetic
acceleration, as expressed in equation \ref{general}, is intended
to represent the dominant term of the magnetic interaction, with
$K$ containing the relevant parameters determining the effective
magnetic time-scale.

Within the above description (\ref{general}) there are still a
number of different possibilities to model the inner disc -
magnetosphere interaction. In this paper we shall use the
diamagnetic blob accretion (DBA) model, although it is possible to
formulate the diametrically opposite case (complete magnetic
penetration of the disc) in the general form (\ref{general}) as
well. Wynn, Leach \& King \shortcite{wyliki} show how to calculate
the appropriate coefficient $K$ for the latter case and we plan to
repeat our calculations in the future, using this prescription.

In the DBA approach one assumes that the fluid constituting the
accretion flow, i.e. the disc and its surroundings, is composed of
blobs immersed in a dilute interblob plasma (see Aly \& Kuijpers
1990). The blobs behave diamagnetically in the presence of the
stellar magnetic field and thus suffer a surface drag force acting
on them. The model has since its introduction been applied to
various systems, notably to the intermediate polar class of
cataclysmic variables (CV). Wynn \& King \shortcite{wyki} and
Wynn, King \& Horne \shortcite{wykiho} incorporated the
diamagnetic drag force into a particle hydrodynamical numerical
code (HYDISC), originally developed by Whitehurst \shortcite{wh}
to simulate the accretion disc in non-magnetic close binary
systems. They found important properties of the white dwarf's spin
evolution and its effect on the disc and explicitly applied their
findings to the moderately magnetic CV AE Aqr. A more recent study
using this approach was its application to the system EX Hya
\cite{kw}.

The DBA approach was extended to the study of accreting T Tauri
stars for the first time by King \& Regev \shortcite{kr},
hereafter KR. They performed calculations of {\em individual} blob
(that is, ballistic) orbits, including the interaction with
magnetic loops of the central star and have shown that this
mechanism can eject blobs from the system, in directions pointing
away from the disc plane. By estimating the angular momentum thus
expelled and including it in the overall angular momentum budget
KR found that a stellar spin equilibrium value is compatible with
the observations (i.e. $\sim$ an order of magnitude less than the
breakup value) if the magnetic loops extend to a few stellar
radii. Pearson \& King \shortcite{peki}, hereafter PK, generalised
the above work into a $N$-particle simulation, in a similar manner
to the one mentioned above for CVs. This work also focused on the
issue of the slow observed T Tauri spin rates and confirmed the
idea that such equilibrium spin rates can be achieved by the
expulsion of matter from the disc until the corotation radius (see
below) coincides with the edge of the magnetic loop. The
additional new finding, not anticipated by KR, of this work was
that the ejection of material comes to a halt as the star
approaches its spin equilibrium value.

The problem of the slow spin rates and rotational evolution of T Tauri
stars as a result of the star's magnetic interaction with its accretion
disc has been also approached, in a number of works, by analytical and
semi-analytical methods, that is, stopping short of multi-dimensional
numerical simulations. K\"onigl (1991), Cameron \& Campbell (1993) and
Armitage \& Clarke (1996) used quite different approaches and all found
that a stellar dipole magnetic field of strength $\la 1~ {\rm kG}$ is able
to regulate the stellar spin to a quasi-static value, in the observed
range. In the process of the magnetic interaction the inner accretion disc
gets disrupted and the inner (abrupt) termination radius of the disc
was estimated in the above papers.

KR and PK used the DBA approach focusing on the
ejection of mass from the disc, in an effort to link the low spin
rates of T Tauri stars with outflows from young stellar objects
(YSO) within a unified scenario. In the present work we utilise
the DBA model in the calculation of the properties of the steady
(or quasi-steady) accretion disc itself, that is, the material
that remains close to the equatorial plane while spiraling in due
to viscous torques, {\em after} the spin period has already
stabilised. Some pecularities of T Tauri {\em light-curves} (i.e.
the temporal luminosity variations) have already been treated
using a model based on single blob ballistic orbits, which leave
the the disc plane but ultimately return and impinge on the disc
surface \cite{ulregber}. In order to investigate whether the DBA
model is consistent with some of the basic properties of T Tauri
{\em spectra} (see Bertout 1989 and references therein) and in
particular the above mentioned IR excess, we have performed
numerical simulations using a code based on HYDISC, suitably
modified so as to adapt it for the problem at hand.

This paper is organised as follows. In the next section we discuss
the basics of blob dynamics and estimate the relevant time-scales.
\S 3 describes the numerical code used in the accretion flow
simulation and the procedure for finding the spectrum emitted by
such flows. In \S 4 the results of our simulations for a score of
parameter values are described in some detail and finally, \S 5
summarises this work in comparison to the results of other
approaches.

\section{Diamagnetic accretion flow dynamics}

Following KR and PK we model the material of a T Tauri disc as being
composed of gas blobs immersed in a tenuous interblob plasma. These
blobs are assumed to behave diamagnetically in the presence of the magnetic
field of the central star, which threads the plasma. Drell, Folley
\& Ruderman (1965) showed that the typical time-scale on which a
diamagnetic object of mass $m$ (i.e. a blob) loses energy, when it moves
in a magnetic field $B$, is
\begin{equation}
   t_{\rm d} = {c_A m \over B^2 l_b^2}
\end{equation}
 where
 $c_A = \sqrt{B^2/(4 \pi \rho_p)}$ is the Alfv\'en speed of the interblob
 plasma, whose density is $\rho_p$, and $l_b$ is the blob size.
 There is a minimum condition on the conductivity for this equation to hold
 and, as it was shown by KR, this condition is satisfied for all reasonable
 parameters in the conditions appropriate for T Tauri accretion.

 This allows to cast the diamagnetic force per unit mass, acting on
 a blob, into one simple formula, whose form is in accordance with
 equation \ref{general}:
 \begin{equation}
 {\bf f}_d = - K_d\, [{\bf u}_b - {\bf u}_f]_\bot\; \hat{\bf n},
 \label{perp}
\end{equation}
where $K_d=1/t_{\rm d}$ (see below) is the diamagnetic drag
coefficient. The magnetic drag force obviously vanishes if the
blob velocity ${\bf u}_b$ {\em relative} to the velocity of the
local field line ${\bf u}_f$ is {\em parallel} to the field lines,
it is proportional to the perpendicular (to the local field)
component of this relative velocity and is directed in the normal
direction $ \hat{\bf n} $ to the field. This is the meaning of the
shorthand vector notation in equation \ref{perp}. Its explicit
component form will be given below.

Assuming that the magnetic field rotates with the star and does
not change on the time-scale of interest, the above equation is
best explicitly written in the frame of reference {\em rotating
with} the central star and had the form
\begin{equation}
{\bf f}_d = - K_d\, [{\bf v} - ({\bf v}\cdot{\bf b}){\bf b}],
\end{equation}
where ${\bf v}$ is the blob velocity {\em in the rotating frame},
that is, relatively to the field lines, and ${\bf b}$ is a unit
vector in the local field direction (see KR).

Substituting explicitly the above form of $c_A$ and $m=l_b^3
\rho_b$, where $\rho_b$ is the blob density, in the drag
coefficient one gets
\begin{equation}
K_d({\bf x}) = 2 \sqrt{\pi}\, \rho_p^{1/2} (\rho_b\, l_b)^{-1}\, B
\equiv K\, f({\bf x}),
\end{equation}
where the second equality parametrizes $K_d$ in terms of a
constant coefficient $K$ and a spatially dependent (${\bf x}$
denotes the position vector) function, which is of order $1$. The
blob parameters and their space dependence, as well as the plasma
density, are rather uncertain. The only variable which can be
explicitly written is the magnetic field (in the approximation
that it is fixed in the rotating frame, that is, anchored to the
star), because it can be chosen a priori.

In this work we shall mainly use a dipolar magnetic field
configuration and group, for the sake of convenience, all the
other variables into a single fixed parameter, incorporating in it
most of our ignorance of the complex plasma instabilities, which
are supposed to create and shape the blobs. Within this framework
we have
\begin{equation}
K_d = K\, (R/R_*)^{-3}\, [1 + 3 (z/R)]^{1/2},
\label{key}
\end{equation}
in cylindrical coordinates $(R,z)$ whose $z$ axis coincides with
the dipole axis. $R_*$ is the stellar radius and
\begin{equation}
K = \sqrt{\pi} \rho_p^{1/2} (\rho_b\, l_b)^{-1} B_0,
\label{key0}
\end{equation}
where $B_0$ is the polar field strength. This $K$ is assumed to be
constant and will serve as a parameter of our problem. For a
discussion of the different options to chose $K_d$ see e.g. Wynn,
King \& Horne (1997).

We focus in this work on the case of an {\em aligned} dipole, that
is, when the dipole axis coincides with the disc axis, but will
briefly discuss (in \S 4.3) the inclined dipole case as well.

The magnetic drag term has to be added to the blob equations of
motion, which include also gravitational, centrifugal and coriolis
terms (in the rotating frame). Scaling the independent variables
by their natural values, that is, lengths by the corotation
radius, $R_{\rm co} \equiv (G M_*/\Omega^2)^{1/3}$ (with $M_*$ the
central star's mass and $\Omega$ its rotational angular velocity)
and time by $1/\Omega$, the following blob equations of motion are
found \cite{ulregber}:

\begin{equation}
\ddot x = 2 y + x - {x \over R^3} - k_d({\bf x})\, [\dot x - (\dot
{\bf x}\cdot {\bf b})\, b_x], \label{xeq}
\end{equation}

\begin{equation}
\ddot y = -2 x + y - {y \over R^3} - k_d({\bf x})\, [\dot y -
(\dot {\bf x}\cdot {\bf b})\, b_y], \label{yeq}
\end{equation}

\begin{equation}
\ddot z = - {z \over R^3} - k_d({\bf x})\, [\dot z - (\dot {\bf
x}\cdot {\bf b})\, b_z], \label{zeq}
\end{equation}

These constitute the nondimensional equations of motion for the
cartesian components, in the {\em rotating frame}, of the blob
trajectory ${\bf x}(t)$ and are thus ready to be incorporated into
HYDISC (see in the next section). $R$, the cylindrical radius, is
obviously $R \equiv (x^2 +y^2)^{1/2}$ and the calculation of ${\bf
b}$ is quite simple for an aligned dipole. For a tilted dipole, a
somewhat more involved (but straight forward) calculation is
needed.

The drag term has been written as $k_d$, the nondimensional
version of $K_d$ (of equation \ref{key}). The explicit form of
this function, following from equations \ref{key} and \ref{key0},
brings out the physical meaning of the constant coefficient $k$
appearing in front of the nondimensional spatially dependent part
of the magnetic field
\begin{equation}
k_d({\bf x}) =  k\, (R/R_*)^{-3}\,[1+3 (z/R)]^{1/2}, \label{kapa}
\end{equation}
which is
\begin{equation}
k = 2.82 \left( {B_0 \over 100\, {\rm G}}\right)\, \left(\rho_p
\over 10^{-10}\, {\rm g~cm^{-3}} \right)^{1/2}\,\left({l_b\, \rho_b
\over 100\, {\rm g~cm^{-2}}}\right)^{-1}\, \left( P_* \over 10^6\, {\rm
s} \right) \sim {t_{\rm rot} \over t_{\rm mag}(R_*,100 {\rm G})}.
\label{kapa0}
\end{equation}

Here reasonable (order of magnitude) estimates of the parameters
for accreting T Tauri systems have been used to scale the relevant
variables. $P_* \sim t_{\rm rot}$ is the stellar spin period and
$B_0$ the polar magnetic field strength. The disc density scaling
value is estimated from models of standard accretion discs around
T Tauri stars, using a representative accretion rates of $\dot M
\sim 5~ 10^{-8} M_\odot {\rm yr}^{-1}$ \cite{bibi}. The blob
parameters are rather uncertain, but it is reasonable that the
blob size is probably of the order of the vertical scale height in
the disc while its density should be several orders of magnitude
larger than that of the disc. This gives the value with which we
scale $l_b\, \rho_b$. In any case, our results will be seen to
depend only on the parameter $k$ (and several other quantities
related to the numerical model) and hence on the particular
combination given in equation \ref{kapa0}.

Time-scale estimates can now be easily seen to demonstrate that
for $k \sim 1$ the magnetic drag time-scale near the star (where
$R\sim R_*$ and thus the contribution of the space dependent part,
arising from eq. \ref{kapa}, is very close to $1$) for a polar
field of $\sim 100 {\rm G}$, denoted here by $t_{\rm mag}(R_*,100
{\rm G})$, is much larger than the dynamical time-scale there,
since the latter is much shorter than $t_{\rm rot}$ (T Tauri stars
are slow rotators). Thus we do not expect a significant effect
near the star due to the magnetic drag in this case. Since the
magnetic field decays (and thus the magnetic drag time scale
grows) faster with $R$ than the dynamical time-scale, the effect
of the magnetic field on the motion will be small everywhere for
$k \sim 1$

Let us examine now what happens at the corotation radius. From the
functional dependence of the dipole field (eq. \ref{kapa}) we have
the term $(R_{\rm co}/R_*)^{-3} =(\Omega/\Omega_K)^2$, with the
ratio of the $\Omega$-s evaluated at the stellar surface. This
term is of the order of $10^{-2}$ in T Tauri stars, but $t_{\rm
rot}$, by definition, is equal to the dynamical time scale at the
corotation radius. If one chooses $k \sim 100$, a significant
effect of the magnetic force at corotation is expected and
obviously more so inside corotation. Thus we expect a significant
modification of the inner accretion disc in this case.

These time-scale estimates provide the rationale for our choice of
the $k$ parameter range (or $B_0$ with all the other parameters
constant) in the simulations (described in \S 4). Note that the
viscous time-scale did not enter our considerations here. In
accretion disc theory it is always assumed that this time-scale is
much larger than the dynamical time-scale and consequently if the
the magnetic time-scale approaches the dynamical one they are both
much shorter than the viscous time scale. However, in regions
where the magnetic time-scale is extremely long (far out in the
disc, or even quite close in, if the field is weak enough) it
might be even longer than the viscous time-scale and thus the
magnetic drag would not interfere with the viscous evolution of
the disc.

\section{The numerical method}

\subsection{General considerations}

To facilitate an accretion flow simulation in our system we adapt
HYDISC, a numerical particle code originally written by Whitehurst
\shortcite{wh} to study accretion in CV, to the problem of a
single central magnetic accretor (a T Tauri star). The
introduction of the magnetic interaction into the particles'
equations of motion, using the DBA approach, has already been
explained in the previous section. Our blobs are being considered
as the particles of the particle scheme, utilized in HYDISC to
simulate the hydrodynamical aspects of the flow, like pressure and
viscosity. As it was already mentioned in the Introduction,
similar schemes based on HYDISC have been designed for simulating
accretion flows in magnetic CV \cite{wyki,wykiho} and also for
investigating the outflows and spin evolution of an accreting T
Tauri star \cite{peki}.

Our calculation is, in principle, similar to the latter work
because the gravitational effects of the companion star are
removed from the code, however the purpose of this work and
therefore the simulation details are quite different. We are
interested in obtaining a steady-state structure of the disc
itself and the spectral shape of the emitted radiation after the
spin period of the central star has settled onto its equilibrium
value. This calls for the determination of an appropriate spatial
computational domain, a suitable particle injection scheme
(initializing the calculation) and the need for a method to assess
the temperature (and from it, the integrated spectrum) of the
accreting matter.

The inner and outer boundaries of our {\em computational region}
are the stellar radius, $R_{*}$, and the "escape" radius, $R_{\rm
out}$ (which is a parameter typically set at $\sim 10\, R_{\rm
co}$), respectively. Particles are injected into the computational
domain (in a manner described below) and if during the course of
the evolution they are found to leave this domain, they are
removed from the calculation altogether. Once in the computational
zone, each particle is subject to the field forces formulated
explicitly in the rotating frame by the equations of motion
\ref{xeq} through \ref{zeq}. In addition, short-range forces,
reflecting the interaction between adjacent particles and thus
simulating the hydrodynamical aspect of the dynamics, are employed
using the HYDISC version of the particle hydrodynamics approach
\cite{wh}. These forces are controlled by three parameters: $C$,
determining the pressure force; $Q$, a constant of order unity and
$r_{\bf v}$, the "chaining" mesh size. Physically $C$ is the sound
speed, $Q$ is in fact the coefficient of restitution of the
particle-particle collisions and is less than $1$ for inelastic
collisions, which must be assumed to simulate viscous flows, and
$r_{\bf v}$ is the viscous scale-length (for details, see
Whitehurst 1988).

The fact that HYDISC separates the treatment of the long-range
(field) forces from the short-range (hydrodynamical) ones makes
this code convenient for application to our problem. Most changes
are introduced in the field part (as explained in the previous
section on the equations of motion) and in the initialization
schemes, that is, particle injection (see below), while the
treatment of particle interactions remains essentially the same as
in HYDISC, with an appropriate choice of the parameters.

The structure of HYDISC, as explained above, allows also to
estimate, relatively easily, the energy dissipation (see below).

\subsection{Particle injection and disc initialization}

Our problem involves an accretion disc around a single star and
thus the injection scheme has to be chosen accordingly. The disc
around a T Tauri star is not being continuously replenished by an
external stream, as is the case in discs fed by the secondary star
in a close binary system, like a CV. This said, we still may chose
between two distinct possibilities: a "burst" injection, in which
all particles are injected simultaneously (or almost so) and a
continuous injection. In both cases we have also to select the
{\em spatial} distribution of the injected particles. Clearly, in
order to obtain meaningful results the ultimate configuration
should not depend on the initial conditions. Computation time
limitations dictate that the computational zone must be limited
(as well as the total number of particles). Thus, the seemingly
best option, i.e. to continuously inject particles at a radius
much larger than $R_{\rm co}$ and wait for the disc to be built up
by viscosity, is prohibitive. We have adopted instead what we call
a "modified burst" injection scheme.

The particles were injected at an equatorial ring around a radius
$R_{\rm inj}$ with a Gaussian spread in the individual particles
injection radii and a small Gaussian displacement out of the disc
plane as well. Each particle was initially assigned an azimuthal
Keplerian velocity plus a random contribution, appropriate for the
local disc temperature. The temperature of the disc, used to
calculate the above mentioned random contributions to the injected
particles' velocities and the sound speed (and thus the constant
$C$), was estimated using the analytical standard disc model. The actual
formula which we have used is identical to formula 6 of PK:
\begin{equation}
T(r)=T(3) \left[ {27 \over 1-3^{-1/2}}\, r^{-3}(1-\sqrt{r}) \right]^{1/4},
\end{equation}
where $r$ is the radius in units of $R_\ast$ and T(3), is the temperature
at 3 stellar radii, where true estimates are available \cite{bibi}.
Although the temperature in our disc is somewhat
different, this estimate actually influences only the pressure
in the disc (in addition to the initial condition)
and therefore the error it introduces should be negligible
(see PK).

In order not to lose too quickly
significant numbers of particles from the simulations, the
"modified burst" scheme consisted of extending the burst duration
to a finite (typically a few stellar rotation periods) time
$\tau_{\rm b}$, with the injection rate decaying with time in a
Gaussian manner. In addition, particles were added at a constant
(but rather slow as compared to the burst) rate $\dot N_{\rm
cont}$ so as to compensate for particles lost from the
computational zone (accreted or expelled). The spatial
distribution of the added particles was kept fixed in time. Thus
the modified burst scheme actually gave rise to the following
number of injected particles as a function of time (expressed here
in units of the rotation period)
\begin{equation}
N_{\rm inj}(t)=N_{\rm b}(t) + \dot N_{\rm cont} \, t,
\label{inject}
\end{equation}
where $N_{\rm b}(t) = \int_0^t \dot N_0 \exp(- t_1^2/2 \tau_{\rm
b}^2)\, dt_1$ . Typically, in our simulations the total number of
particles in the burst (i.e. $N_{\rm b}(t=3 \tau_b)$ say) was
between 5000 and 10000 and $\dot N_{\rm cont} = 150$ particles per
rotational period (see below).

\subsection{Calculation of the emitted spectrum}

With the present DBA scheme it is impossible to obtain a detailed
thermal structure of the accretion flow and therefore we aim at
approximating just the effect of the magnetic field on the emitted
spectrum, {\it relatively} to the non-magnetic ($k=0$) case. We
assume that the disc is optically thick in the region of interest
and that the energy dissipated in it by the viscous interaction is
immediately locally radiated out in the vertical direction. These
assumptions are essentially the same as in the standard Shakura \&
Sunyaev treatment of geometrically thin and optically thick
accretion discs. What is different here is the fact that the
viscous dissipation rate must be computed from the details of the
numerical particle scheme and can not be found directly from the
disc's viscous flow. As explained above, a part of the kinetic
energy of the particles (blobs) is lost in collisions as these are
inelastic ($Q<1$). After dividing the computational zone into
cells, by a two-dimensional grid in the disc plane, one can
calculate the energy released in each cell $\Delta E_{ij}$ during
a given time $\Delta t$, provided the {\em masses} of the
particles are known. The composite spectrum can then be found by
adding up the contributions of all the cells in the computational
domain, with each cell assumed to contribute a black-body emission
spectrum, corresponding to its effective temperature
\begin{equation}
T_{ij}^{\rm eff}= \left({\Delta E_{ij} \over \Delta t A_{ij}
\sigma}\right)^{1/4},
\end{equation}
where $A_{ij}$ is the cell surface area and $\sigma$ is the Stefan
radiation constant.

As explained in \S 2 (see equations \ref{kapa} and \ref{kapa0}) we
have lumped the unknown values of the blob properties into a
single parameter and chosen a specific spatial dependence for it.
Thus for a given value of the coefficient $k$, different
combinations of the constituent parameters (and among them the
blob's mass) is allowed. Assuming some fixed average blob mass can
thus give a spectral shape, which should be considered as a {\em
qualitative} estimate and have a {\em relative} meaning, that is,
the spectral shapes for different values of $k$ can be perceived
as an assessment of the effect of the magnetic field strength on
the shape of the spectrum. In addition, since our computational
zone does not include (due to computing time limitations) the
extended outer portion of the disc, the results must be
interpreted remembering this fact. In particular, since the outer
cool parts of accretion discs contribute mainly to the long
wavelength part of the emitted spectrum, our computed spectra
clearly {\em underestimate} the spectrum strength for large
$\lambda$ (see below in Fig. 3).

\section{The simulations}

\subsection{Simulation parameters}
In choosing the parameters of the simulations we have tried to
optimize the sometimes conflicting demands to obtain, on the one
hand, meaningful results and on the other hand to limit, as much
as possible, the CPU demands of the numerical calculations.

We have set the stellar parameters in all the simulations to be
$M_*=M_{\sun}$, $R_*=1.5 R_{\sun}$, $P_*=10^6\, {\rm s} \sim 11\,
{\rm days}$, giving for the corotation radius $R_{\rm co} \approx
1.5\; 10^{12} {\rm cm} \sim 14 R_*$. As explained before, the
corotation radius serves as our length unit.

Experimenting with different injection radii (the location of the
centre of the injection ring), we have found that with the
injection scheme (\ref{inject}) an injection radius of $R_{\rm
inj}=1.4 R_{\rm co}$ is effective in building up rather quickly an
accretion disc, which does not differ essentially from the one
resulting (after a longer calculation) from a larger injection
radius. In these tests we have shut off the magnetic interaction,
but remembered from previous works that the injection radius
should not be too close to corotation. Consequently in all the
simulations we have used this value of the injection radius, with
a Gaussian spread of up to $\pm 0.1 R_{\rm co}$.

As discussed in the previous section, the parameters related to
the particle scheme treatment of the pressure and viscosity
include $C$, $Q$ and $r_{\rm v}$. We chose $C$ to be equal to the
isothermal sound speed, derived from the standard Shakura \&
Sunyaev disc temperature distribution. $Q$ and $r_{\rm v}$
determine the viscous effects and have to be chosen with care.
$Q$, the restitution coefficient has a global effect on the
simulation and since its reasonable values should lie between
$0.5$ and $1$ ($Q<0.5$ allows particle interpenetration and $Q=1$
corresponds to elastic collisions). We have found that changing
$Q$ within the above mentioned range has only a small effect on
the final quasi steady state, but it influences the length of the
time until such a state is reached. Whitehurst (1988) reached a
similar conclusion by experimenting with $Q=0.6$ and $0.7$ values
in the original HYDISC code. Thus, wishing to shorten the
simulation time, without, however, making the viscosity
unphysically large, we have opted for $Q=0.7$ and used this value
in all our simulations, save for a few test cases.

The parameter $r_{\rm v}$, setting the viscous interaction length
scale, has a local effect on the simulation. When deciding on the
range of variation for this parameter we have taken into account
the fact that too small a value of $r_{\rm v}$ causes the viscous
evolution time to be unacceptably long. In addition, because the
number of cells in the computational range becomes large, one has
to increase the number of particles in order to keep the particle
treatment meaningful. Both effects increase the CPU time and we
have found that for $r_{\rm v} \la 0.02$ we were unable to
efficiently compute the disc evolution. On the other hand for
$r_{\rm v}$ approaching the value of $0.1$ the increase in
viscosity causes the particles to clump together and the evolution
becomes unrealistically viscosity dominated. Performing test
calculations with no magnetic interaction ($k=0$) we found that an
initial ring injected around $1.4 R_{\rm co}$ viscously spreads to
a structure very similar to a standard thin disc in a time of
$\sim 100 - 300$ rotational periods when $r_{\rm v}$ is kept
within the range $0.08 \ge r_{\rm v} \ge 0.03$. Using such a
calculation it is also possible to find what is the effective
$\alpha$ (the viscosity coefficient in the Shakura \& Sunyaev
prescription) with a given choice of $Q$ and $r_{\rm v}$. A test
calculation with $ r_{\rm v} = 0.05$ (in units of $R_{\rm co}$,
see below) and $Q=0.7$ was very similar to the evolution of a ring
of matter to as it is found from a standard exercise with an
$\alpha$ viscosity (see e.g. Pringle 1981) value of $\sim 0.1$.

The parameters $k$ and $r_{\rm v}$ were varied in the different
simulations. The range of the variation of $k$ was chosen so as to
correspond to polar stellar magnetic value of up to several
kiloGauss (when the other parameters in $k$ are set to their
representative values, see equation \ref{key0}) and $r_{\rm v}$
was chosen within the range mentioned above.

\subsection{Results of representative cases}

We have performed 12 full simulations of an accretion disc in an
aligned dipole magnetic field for different choices of the pair of
parameters $k$ and $r_{\rm v}$. The simulations were initialized
using the injection scheme described in equation \ref{inject} with
parameters chosen so as to insure that the number of particles
does not drop below $\sim 10^4$ during the simulation. The
simulations were continued until a quasi-steady configuration was
achieved, that is, no significant changes of the structure were
observed over a significant time (see below). We have chosen to
describe here in some detail the simulations with $r_{\rm v}=0.05$
(simulation group 3 in our list), that is, with the viscosity
strength chosen in the middle of its feasible range. As discussed
in the previous section this choice corresponds to an effective
$\alpha \sim 0.1$. Three runs were performed for different values
of the magnetic interaction strength ($k=0,10,100$), where the
non-magnetic case ($k=0$) was run to serve as reference, in
particular for the spectra, which have quantitative meaning only
in comparison to each other (as explained before).

These three runs were all allowed to continue for a total
simulation time of $250 P_*$. A quasi-steady state was achieved
already after approximately half of this time. The CPU time
requirement for each run, with the number of particles kept at
$\sim 10^4$ (the accreted and ejected particles were replenished
as explained above), was about $100$ hours on a Sun ULTRA 2
workstation, when the program was coded in C, to increase its
efficiency. In Fig.~1 we show the surface number density of the
particles (a quantity proportional to the disc surface mass
density, customarily called $\Sigma$) as a function of radius for
the three values of $k$. Note that in the innermost and outermost
parts of the disc, regions in which the surface density curves are
dashed, the number of particles is too small to render the result
to be quantitatively reliable. We also plot on this graph the
analytical result, arising from a standard thin disc formula with
inner cutoff (see below). This is represented by the fully dashed
curve, which which follows the $k=100$ case for most of the range.

\begin{figure*}
 \resizebox{\hsize}{!}{\includegraphics{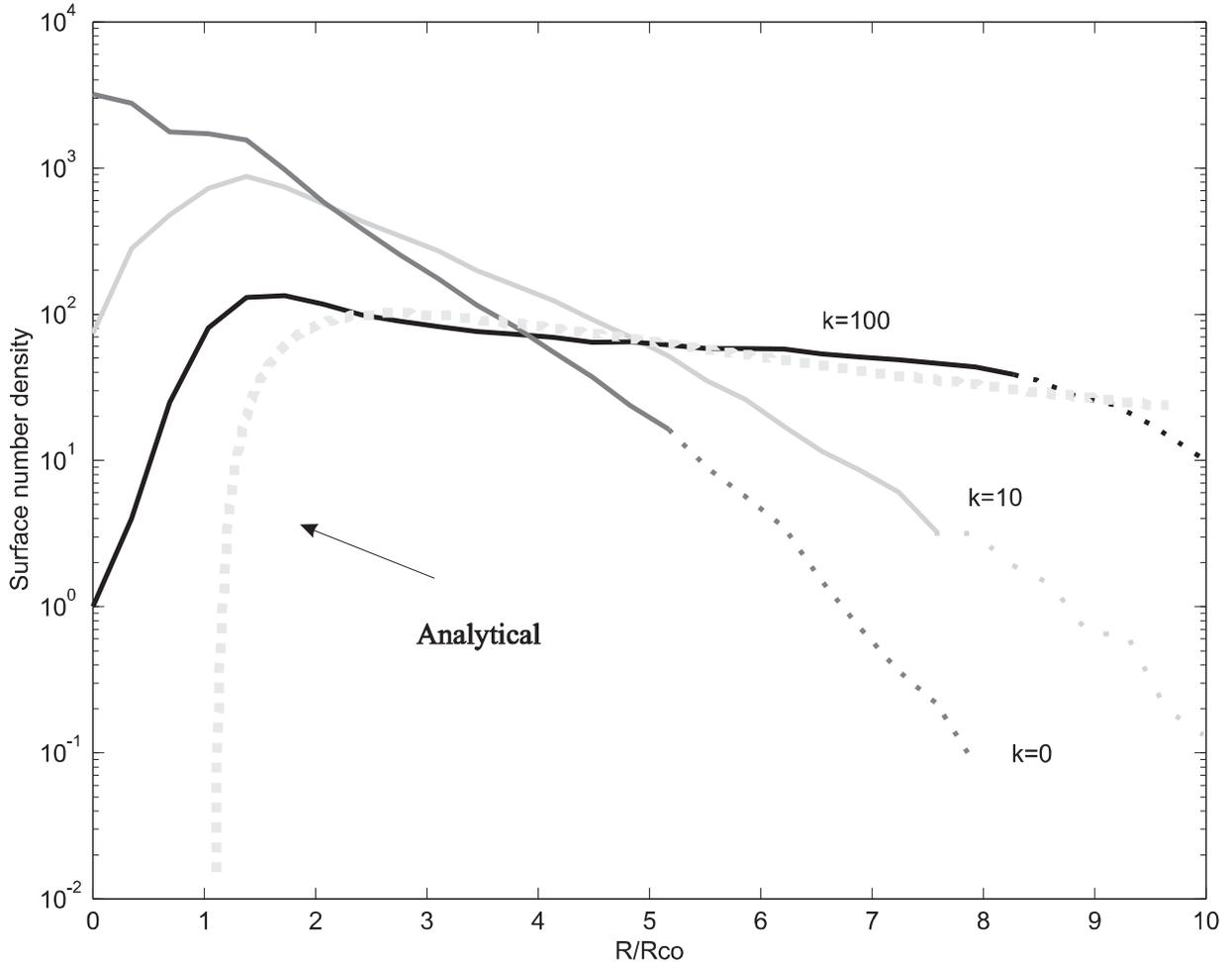}}
 \caption{Surface number density as a function of radius (in
  units of $R_{\rm co}$) for $r_{\rm v}=0.05$ (simulation 3) and
  three values of $k$. The number of particles in the innermost and outermost
  regions, denoted by dashed portions of the curves, is too small to render the
results to be quantitatively reliable there. The fully dashed
curve is an analytical result of a disc truncated at $R_{\rm co}$}
 \label{fig1}
\end{figure*}

From this figure it is evident that the effect of the magnetic
field on the disc structure is important, as it significantly
modifies the functional dependence of the surface density on the
radius. For $k=0$ (no magnetic field) the particles form a rather
"collapsed" with its density growing sharply towards the central
object. This is so because our simulations were tailored for the
high $k$ case and therefore the particle injection rate was too
small for the buildup of a steady disc in a small $k$ case. A
growing magnetic field strength has a dynamical effect on the
disc. As could be expected from a previous study, in which
individual orbits of particles were calculated \cite{ulregber},
and from the DBA simulations of PK, the magnetic interaction drags
blobs residing inside corotation towards the star and pushes out
blobs whose orbits are outside corotation. These effects are
modified here by the fact that the dipole magnetic field decays
outwards and by the interblob interactions. The combined effect is
a significant depletion of blobs close to the star (a "hole") and
an increase of density at large radii (a "flattening" of the
functional dependence of the surface density on radius). The size
and depth of the hole and the extent of the flattening of the
surface density obviously depend on $k$, That is, on a combination
of the disc density (which depends on the accretion rate) and the
strength of the magnetic field.

Analytic estimates of the hole radii at spin equilibrium, as given
by Cameron and Campbell (1993), predict that the disc should end
abruptly close to the corotation radius (but depending also on the
fastness parameter of the star). Our numerical results do not
yield an abrupt disc termination, but the value of $\sim R_{\rm
co}$ is certainly consistent with these results. In the following
comparisons of our results to analytical solutions we have used
this value of the inner cutoff radius, that is, the analytical
formulae giving rise to the fully dashed curves in Figs. 1 and 2
(see below) result from
\begin{equation}
\Sigma(R)={\dot M \over 3 \pi \nu} \left[1-\sqrt{R_{\rm
co}/R}\right] \label{ss1}
\end{equation}
and
\begin{equation}
T^4(R)={3 G M \dot M \over 8 \pi R^3 \sigma } \left[1-\sqrt{R_{\rm
co}/R} \right], \label{ss2}
\end{equation}
with the values of $\nu$ and $\dot M$ chosen appropriately, from
our calculation, so as to enable the comparison.

\begin{figure*}
 \resizebox{\hsize}{!}{\includegraphics{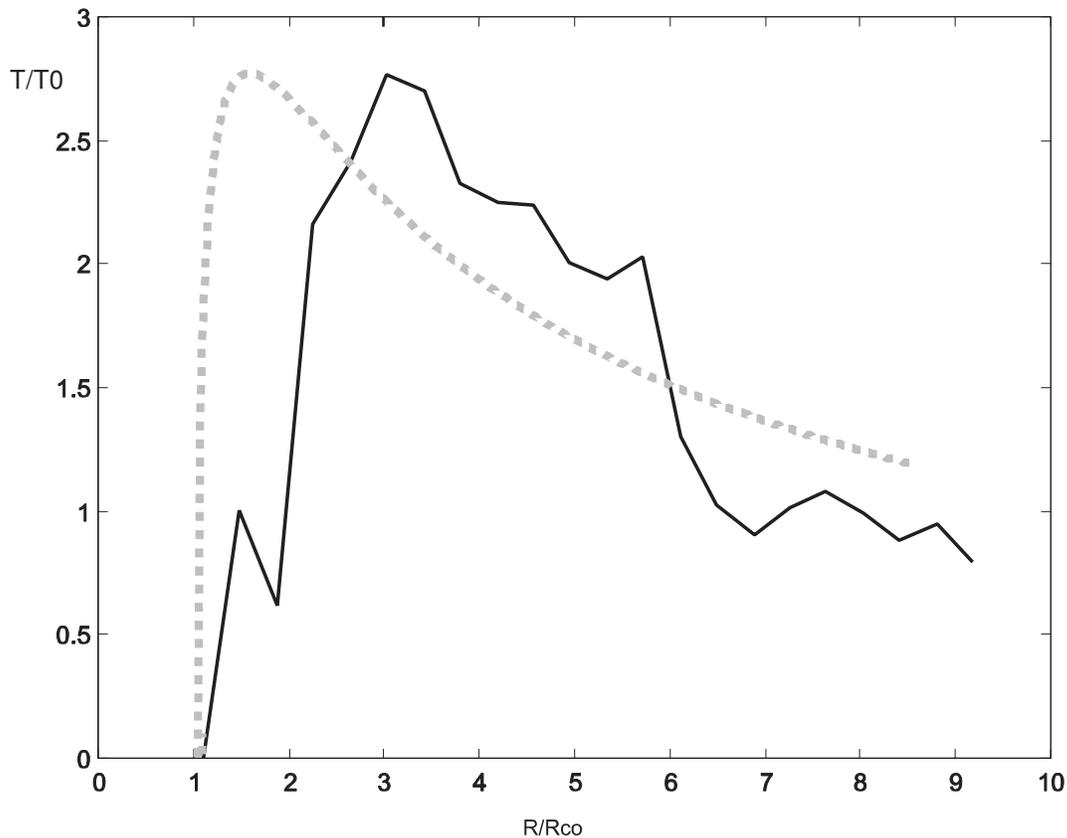}}
 \caption{The effective temperature (in arbitrary units) as a function of radius
 (in units of $R_{\rm co}$) for $r_{\rm v}=0.05$ (simulation 3) and
   $k=100$. The dashed curve is the analytically obtained temperature
   distribution of a classical accretion disc, cut off inside at $R_{\rm in} = R_{\rm
   co}$ (see \ref{ss2})}
 \label{fig2}
\end{figure*}

If Fig.~ 2 we show the comparison of the analytical (effective)
temperature distribution as given in (\ref{ss2}) with the
numerical result for the case of $k=100$, when the outer parts of
the disc resemble most the analytical result (see Fig.~1). The
lack of the inner, hottest, part in the numerical results is
apparent. In these calculations the method described in \S 3.3 was
used to monitor the energy dissipation during the simulations and
calculate the discs temperature structures and their spectra. The
spectral flux distributions ($\lambda F_{\lambda}$) in the three
cases of $k$ depicted in Fig.~ 1 are displayed in Fig.~3.

It is apparent from the figure that with increasing $k$ the
spectral distribution of the emitted radiation is shifted towards
longer wavelengths. This trend is quite clear since as we have
seen in Fig.~1 and 2, the hottest material, which resides close to
the star, is significantly depleted by the magnetic interaction.
In addition, as a result of the increase of the surface density in
the outer cooler parts, the contribution to the long-wavelength
part of the spectrum is more prominent. We stress once again that
the temperature calculation and hence that of the spectrum suffers
from the fact that the {\em mass} of the blobs is essentially
unknown, so that the results displayed in Fig.~3 are of a
qualitative nature and their value is mainly in uncovering the
trend of the spectral behaviour. In addition, computer time
limitations do not allow us to extend the calculation to the outer
parts of the disc (here, in simulation 3 we have $R_{\rm out}=10
R_{\rm co}$). Thus the spectral energy distributions in Fig.~3
fail to account for the cool material lying outside the outer
limit of the computational zone and which should have an effect of
increasing the strength at long wavelengths. Since the disc is
more extended for large $k$ this effect should matter the most in
the strongly magnetic case. The actual structure of the accretion
disc for the three values of $k$ can be seen in Fig.~4, in which
maps of the the surface density and temperature are shown.

\begin{figure*}
 \resizebox{\hsize}{!}{\includegraphics{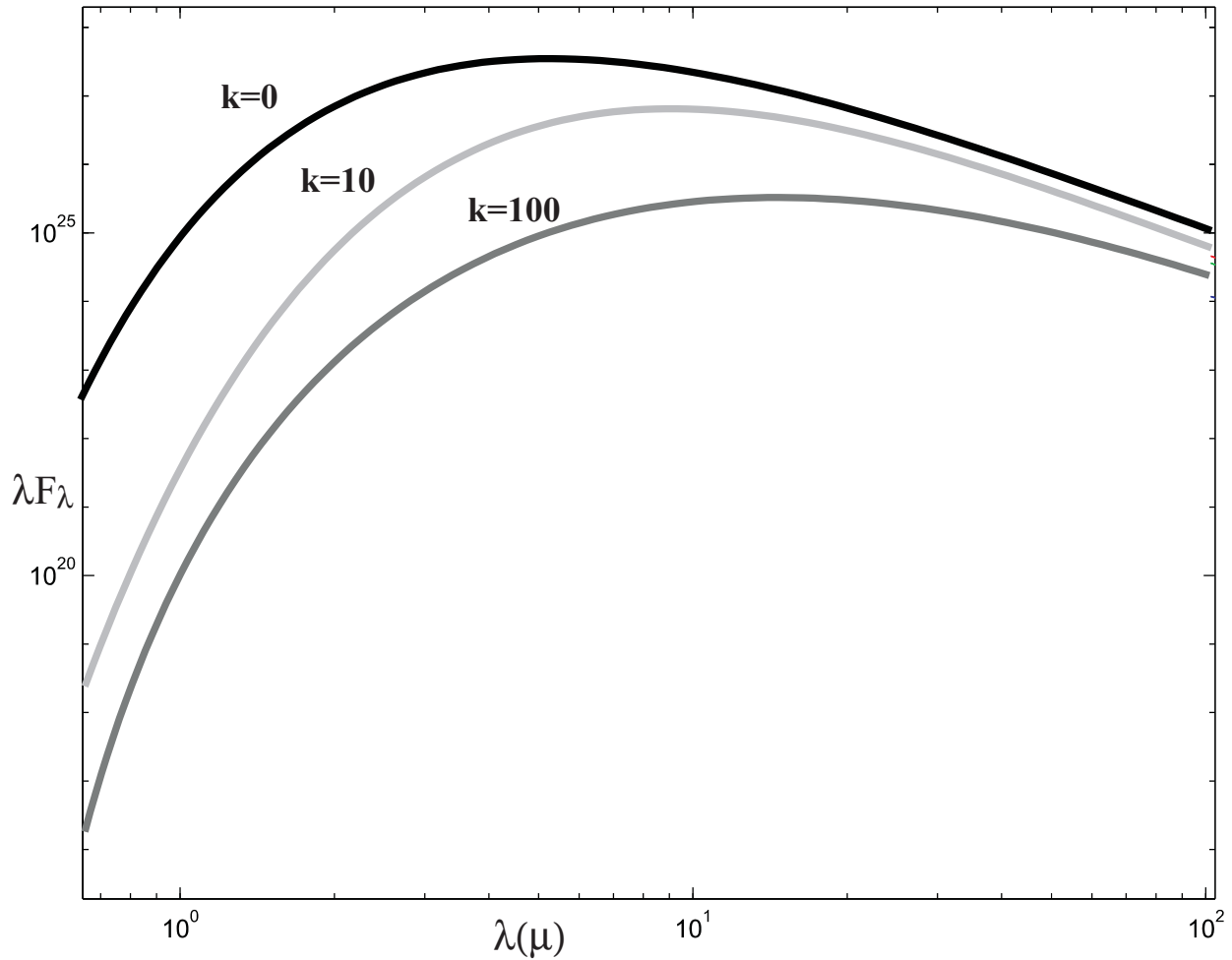}}
 \caption{Spectra of the accretion discs in simulation 3. $\lambda
F(\lambda)$ in arbitrary units (see text) is plotted as a function
of wavelength $\lambda$ in microns for three values of the
magnetic interaction parameter $k$.}
 \label{fig3}
\end{figure*}

From the examination of Fig.~3 and the above considerations it is
thus reasonable to conclude that the magnetic interaction, at
least as it is dealt with in this work, tends to flatten the
spectrum for long wavelengths, consistently with the observed IR
excess mentioned in the Introduction.

\begin{figure*}
 \resizebox{\hsize}{21.cm} {\includegraphics{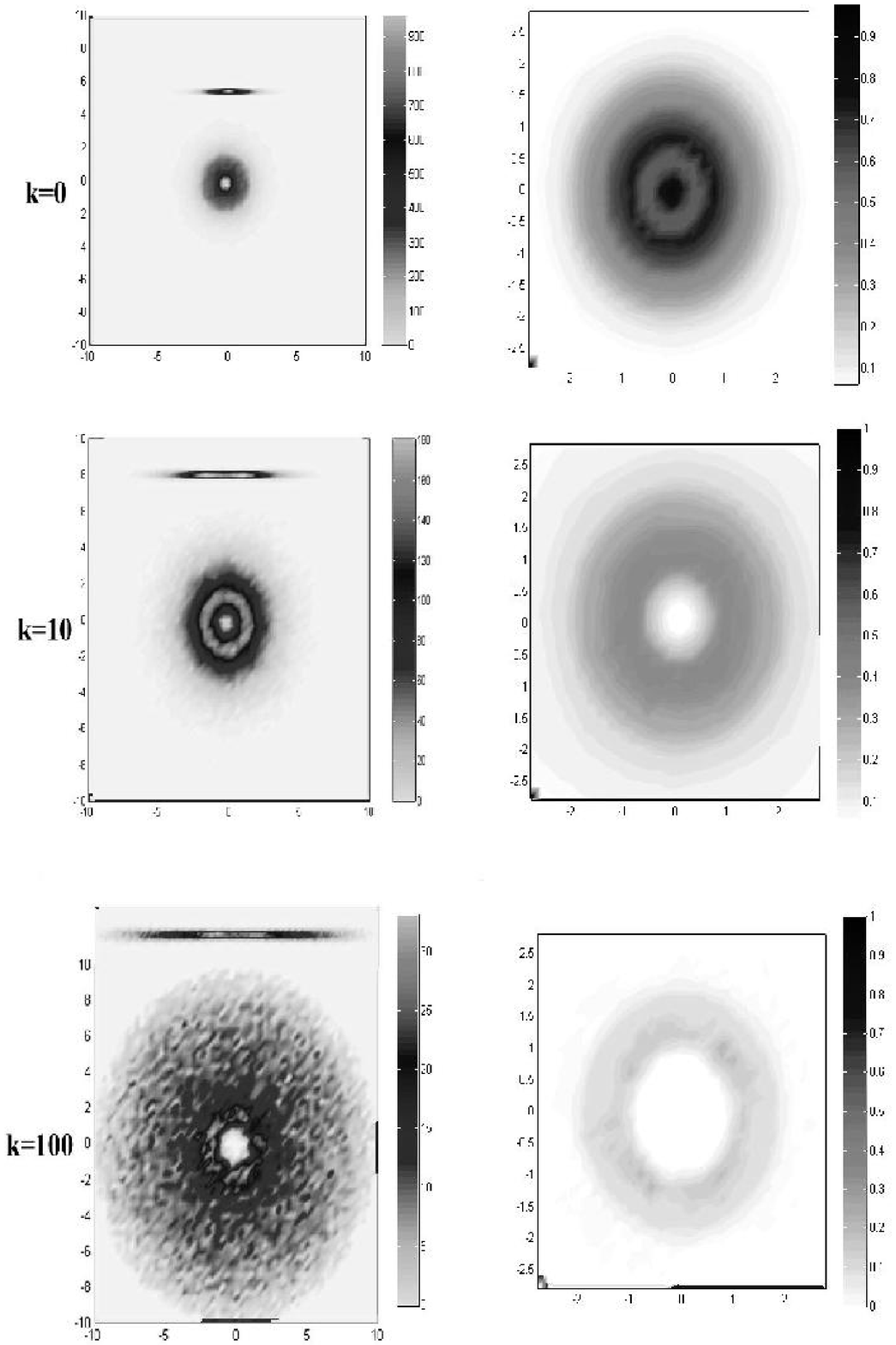}}
 \caption{Surface density (left panels) and temperature (right panels) maps
for simulation 3 ($r_{\rm v}=0.05$) for the three values of the
magnetic interaction strength. Arbitrary units are used and in the
surface density maps a side view of the disc is also given, in the
upper part of each panel.Surface density (left panels) and
temperature (right panels) maps for simulation 3 ($r_{\rm
v}=0.05$) for the three values of the magnetic interaction
strength. Arbitrary units are used and in the surface density maps
a side view of the disc is also given, in the upper part of each
panel.}
 \label{fig4}
\end{figure*}

The results of all the other simulations (i.e. the ones with a
different value of the viscosity scale parameter $r_{\rm v}$ are
similar to the ones described above and thus the conclusions are
similar. The actual value of $r_{\rm v}$ has only a little effect
on the final (i.e. quasi-steady) surface density and temperature
distribution, as long as this value is within the calculation
feasibility range (see above). The difference is in the time that
has to pass until the configuration settles down. This is
understandable as the evolution time is essentially the viscous
time scale and the latter decreases with increasing $r_{\rm v}$.

\subsection{Effects of special magnetic field geometry}

We have performed, in addition to the simulations described above
in which the magnetic field was that of an aligned dipole, two
similations with different magnetic configurations. In the first
one we have used a magnetic dipole whose axis was tilted to the
disc axis and in the second one a "dipole slice" (see below) was
introduced in order to estimate the effects of a localised
magnetic field configuration.

\subsubsection{Tilted dipole}

A single simulation was performed with $r_{\rm v}=0.05;\,k=100$,
that is, similar to one of the simulations described above, but
with the magnetic dipole axis tilted at an angle of $45^\circ$ to
the disc axis. The results show relatively few changes as compared
to the aligned dipole case. There was almost no perturbation of
the disc matter, relatively to the aligned dipole case, far out of
the corotations radius. This is understandable since the blobs
orbiting at these positions experience in this case a magnetic
field whose direction oscillates with the period of the stellar
rotation (which is considerably shorter than the blob's Keplerian
period). Thus, it is expected that the blobs actually react to the
average magnetic field, i.e. behave in a similar way as in the
aligned dipole case.

This expectation is wrong for blobs close to and inside corotation.
Indeed, their motion should be channeled there by the magnetic
field. This effect could not be observed in our simulations since
the density of the particles in the magnetosphere, above the disc
plane, is very small. A very high number of particles or
alternatively an extremely strong magnetic field is needed in
order to resolve, in our type of calculation, the particle flow
along field lines. This remark obviously applies to the aligned
dipole case as well.

\subsubsection{Dipole slice}

To simulate the case in which the actual stellar magnetic field
resembles a localized loop, similar to the case studied by PK,
we have used a "slice" of the aligned
dipole, with Gaussian fall-off in strength in both the radial
coordinate $R$ and the azimuthal angle $\theta$. Thus we have
performed a simulation with $r_{\rm v}=0.05;\,k=500$ and with the
magnetic field centered on the position $R_0,\theta_0$
\begin{equation}
B_{\rm slice} = B_{\rm dipole}
\exp[-(\theta-\theta_0)^2/2d_\theta^2] \exp[-(R-R_0)^2/2d_r^2],
\end{equation}
where $B_{\rm dipole}$ is a dipole field (see the functional
dependence in equation \ref{key} and $d_R$, $d_\theta$ are the
radial and angular Gaussian fall-off scales in the radial and
azimuthal directions. In the simulation we have chosen $R_0 \sim
R_{\rm co}$, consistently with the findings of KR that this should
be the equilibrium configuration.

The results clearly depend on $d_R$ and $d_\theta$. When we chose
$d_R \sim R_{\rm co}$, the disc beyond $R_{\rm co}$ resembled the
non-magnetic case ($k=0$), since the field is practically zero
there. Closer to corotation and inside it, the density
distribution looked like that of a full but {\em weaker} dipole,
since the dipole slice field effectively acts on the orbiting
blobs only along a fraction $\sim d_{\theta}/2 \pi$ of the blob's
orbit.

\section{Summary and conclusions}

The most prominent feature of our simulation results is the
formation of a low density region (a hole) in the inner part of
the disc as a result of the magnetic interaction. In addition, the
density distribution becomes more spread out, as the slope of
$\log \Sigma(R)$ decreases with increasing $k$. This behaviour is
typical and essentially qualitatively independent of all the other
parameters. Although the "hole" is the more noticeable feature
(see e.g. the density maps in Fig.~4), it is less significant to
overall observable features than the global change in the density
distribution.

We have also seen, by comparing our calculations with what can be
expected from analytical results, that we uncover some features
(like the absence of hot material close to the inner hole) which
can not be obtained from just classical disc models with a hole
inside.

Due to computer power restrictions we were unable to achieve a
high enough resolution, so as to see the accretion flow along the
field lines and outflows from the system. In addition we were able
to simulate only a limited portion of the disc and thus the
calculations of the spectrum reveal only the general trend. This
trend, resulting from the flattening of the surface density
distribution, was always to shift emission power toward longer
wavelengths. The flattening of $\lambda F_{\lambda}$ is most
apparent near its maximum and it is reasonable that were it not
for the close (computational) cutoff in $R$, the shape would
remain flat to longer wavelengths. We propose therefore that the
IR excess in T Tauri stars can be attributed to magnetic
interaction, which modifies the functional dependence of the
surface density in the surrounding discs.

Existing models attempting to explain the IR excess in T Tauri
stars fall into two distinct categories: those invoking
geometrical factors and others, proposing energy dissipation
mechanisms operating preferably in the outer parts of the disc.
Our model suggests a correlation between the spectrum flatness and
the magnetic field strength, appears to be quite robust
(practically any shape of the magnetic field would do) and does
not require assumptions about flared shapes of discs or unusual
energy dissipation modes. At this stage, the results of our
calculation provide little more than support for a {\em
qualitative} promising idea. As it was mentioned in the
Introduction, it is possible to apply a similar prescription to
the case in which the magnetic field penetrates the disc. We can
reasonably expect that the results of such a calculation should
not be too different, at least qualitatively, from the ones
presented in this paper. It appears that all that is required is
magnetic field lines imparting a torque on the gas, which changes
in sign as we cross the corotation radius.

Significant results for both models of the magnetic interaction
can be achieved in high resolution (significantly larger number of
particles). However, to extend the idea into a reliable
quantitative model, full multidimensional MHD simulations,
including radiative transfer, have to be ultimately performed.

\section*{Acknowledgments}

We thank Giora Shaviv for helpful remarks and suggestions. Yigal
Ultchin acknowledges the hospitality of the Astronomy Group at
Leicester University and the financial support extended to him at
the Technion. This research was supported by grants from the
Israel Science Foundation, the Helen \& Robert Asher Research Fund
and by the Fund for the Promotion of Research at the Technion. We
would also like to thank the referee, whose insightful comments
helped considerably in improving this paper.

 \label{lastpage}
\end{document}